\documentclass[twocolumn,epjc3]{svjour3}  

\usepackage{bm}
\usepackage{mathrsfs}
\usepackage{xcolor,color,graphicx,graphics}
\usepackage[all]{xy}
\usepackage{epsfig,subfigure}
\usepackage{latexsym,amssymb,amsmath,amsfonts} 
\usepackage[english]{babel} 
\usepackage[OT1]{fontenc}
\usepackage{makeidx}
\usepackage{hyperref}
\usepackage{color,graphicx,graphics,wrapfig,epsf}
\usepackage{float}
\restylefloat{table}
\usepackage[colorinlistoftodos]{todonotes}
\usepackage{blindtext, xcolor}
\usepackage{verbatim}
\definecolor{red}{rgb}{1,0,0}

\def\+{^\dagger}

\def\<{\leftarrow}
\def\>{\rightarrow}

\def\({\left(}
\def\){\right)}

\def\sc{\mathop{\rm sc}\nolimits}

\newcommand{\bi}{\begin{itemize}} 				\newcommand{\ei}{\end{itemize}}
\newcommand{\benu}{\begin{enumerate}} 		\newcommand{\enu}{\end{enumerate}}
\newcommand{\bd}{\begin{dinglist}{0}}     \newcommand{\ed}{\end{dinglist}}
\newcommand{\bfig}{\begin{figure}[htbp]}  \newcommand{\efig}{\end{figure}}
\newcommand{\bc}{\begin{center}} 				  \newcommand{\ec}{\end{center}}
\newcommand{\be}{\begin{equation}} 				\newcommand{\ee}{\end{equation}}
\newcommand{\bsub}{\begin{subequations}}  \newcommand{\esub}{\end{subequations}}
\newcommand{\ben}{\begin{eqnarray}} 			\newcommand{\een}{\end{eqnarray}}
\newcommand{\ba}[1]{\begin{array}{#1}} 		\newcommand{\ea}{\end{array}}
\newcommand{\bea}{\begin{equation}\begin{array}{rcl}}
\newcommand{\eea}{\end{array}\end{equation}}

\RequirePackage{fix-cm}
\smartqed  
\RequirePackage{graphicx}
\journalname{Eur. Phys. J. C}
\begin{document}

\title{Early evolution of fully convective stars in scalar-tensor gravity
}

\author{D{\'e}bora Aguiar Gomes\thanksref{e1,addr1}
        \and
       Aneta Wojnar\thanksref{e2,addr2} 
}

\thankstext{e1}{e-mail: deboragomes@fisica.ufc.br}
\thankstext{e2}{e-mail: awojnar@ucm.es}


\institute{Laboratory of Theoretical Physics, Institute of Physics, University of Tartu,\\
W. Ostwaldi 1, 50411 Tartu, Estonia \label{addr1}
           \and
           Departamento de F\'isica Te\'orica \& IPARCOS, Universidad Complutense de Madrid, E-28040, 
Madrid, Spain \label{addr2}
}

\date{Received: date / Accepted: date}

\maketitle

\begin{abstract}
In this work, the early evolution of low-mass fully convective stars is studied in the context of DHOST (degenerate higher
order scalar-tensor) theories of gravity. Although it is known that the hydrostatic equilibrium equation is modified for scalar-tensor gravity, the consequent modifications to the early evolution phases of a star were not explored in this framework. With this in mind, we consider three evolutionary phases --- contraction to the main sequence, lithium burning and entrance to the main sequence --- and investigate how each of these phases is affected by the theory's parameter. Taking these effects into account, we are able to show, among other things, that the Hayashi tracks are shifted and the star's age is considerably modified.
\end{abstract}

\section{Introduction}

As a theory that has successfully passed several tests, it is safe to say that General Relativity (GR) is currently our best description of the gravitational interaction \cite{ein1,ein2,will}. However, these successes did not prevent attempts of modifying gravity \cite{Copeland:2006wr,Nojiri:2006ri,nojiri2,nojiri3,Capozziello:2007ec,Carroll:2004de,cantata} which were mainly motivated by the discovery of the accelerated expansion of the universe \cite{hut}. Thus, a modified theory of gravity should stand as an alternative candidate to dark energy models on large scales and, at the same time,  reproduce GR's predictions in Solar System scales \cite{capo,capo2008,so2,clifton2012,so7,so3,hu2007,capo2011}. 

Many approaches can be taken in order to modify gravity. One of them consists in adding extra fields to the theory, such as a scalar field (see e.g. \cite{dicke,fuj}, for more general fields, see \cite{moffat}). When working with scalar-tensor theories that modify GR, usually one considers only those theories that have up to second order field equations as higher derivatives commonly lead to Ostrogradsky instabilities \cite{Ost}. The most general theory obeying this requirement is the so-called Horndeski theory \cite{Horndeski}. However, theories with higher-order derivatives which have degenerate Lagrangians can be shown to be ghost free, giving rise to the Beyond Horndeski class of theories \cite{BeyondH,BeyondH2}, referred further to DHOST (degenerate higher order scalar-tensor) theories.
As already mentioned, the modifications to GR should agree with standard solar system tests. However, in such modifications, the gauge symmetry of GR is usually broken, which generates new degrees of freedom. For instance, in scalar-tensor (ST) theories where the scalar field is coupled to the Ricci scalar, the presence of a fifth force requires the application of a screening mechanism in order to suppress its effects at small scales \cite{screening,screening2,screening3,sunny,shei2}.

Apart from cosmology, which was extensively studied in the framework of Horndeski and beyond \cite{kobayashi,shei1,shei3}, the theory was also studied in the context of relativistic \cite{Astashenok:2021peo,Astashenok:2020qds,Astashenok:2021xpm,Astashenok:2021btj,Odintsov:2021qbq,Odintsov:2021nqa,Oikonomou:2021iid} non-relativistic stars and substellar objects. From the finding that the screening mechanism is partially broken in Main Sequence stars \cite{broken}, the ST theories have joined the class of gravity models  \cite{Saito:2015fza,olekinv,olmo_ricci,review,anetarev} which modify the Poisson and hydrostatic equilibrium equations, which turn out not only to have a non-negligible effects on the inner structure \cite{olek,olek2,olek3} and evolution \cite{aneta2,chow,merce,straight,maria,kart,anetaJup2} of stellar and substellar objects, but this fact also provides tools to constrain those models \cite{saltas1,saltas}. It has been showed that modified gravity alters mass limits such us minimal masses for hydrogen and deuterium burning \cite{sak1,sak2,Crisostomi:2019yfo,gonzalo,rosyadi}, Jeans and opacity masses  \cite{capJeans,anetaJup}, or Chandrasekhar one for white dwarf stars \cite{Chandra,Jain:2015edg,Banerjee:2017uwz,Wojnar:2020wd,Belfaqih:2021jvu,Saltas:2018mxc,kalita,kalita2}. Furthermore, light elements' abundances in the stellar atmosphere seem to be also affected when other than Newtonian model of gravity is applied \cite{aneta3}.  

In the presented work we will follow the steps undertaken previously in \cite{aneta2,aneta3} to examine the early phases of a young star which is contracting to the Main Sequence (MS). Mainly, we will focus on the pre-Main Sequence (PMS) tracks which are given by the effective temperature-luminosity relation, called Hayashi tracks \cite{hayashi}. In this period of time the PMS star possesses sufficient conditions in its core in order to start lithium ignition. Since the lithium abundance is a time-dependent quantity, it will allow us to find the age of an object which just entered the MS phase. Moreover, depending on its mass, such a star can leave its Hayashi track and develop a radiative core, which have a further consequence - it will follow the so-called Henyey track \cite{hen1,hen2,hen3} instead (see the picture in \cite{anetarev} for the early evolution's phases). However, since this feature also depends on the gravity model applied, the maximal mass of a fully convective star on the MS may be different than in the common models based on Newtonian gravity.

Those phases of the low-mass star's evolution were not studied in the context of ST theories. Keeping in mind that in the nearest future we will be supplied with more accurate data from different missions such as e.g. James Webb Space Telescope or Nancy
Grace Roman Space Telescope \cite{vision,voyage,webb,nancy,tess,spitzer,nn}, therefore one should be ready to have the most popular theories of gravity prepared for the data release.  This will allow to use statistical methods, as one will be equipped with rich data samples, to understand them and put constrains on such gravitational proposals with a great statistical power. Since DHOST theories do modify internal properties and structure of stellar objects, the presented findings can be used to further constrain ST theories.  
The low mass stars in globular clusters have been already used to test Standard Model of particle physics and the dark matter candidates \cite{raf2,raf3}; it is expected that when re-analyzing the nuclear processes in the framework of modified gravity, one will be also able to bound some models, as it was done in the case of the dark matter ones \cite{raf1}. It is also a significant fact in favor of studying these objects that the low mass stars have long lifetimes and hence even very small effects as presented in this work, can accumulate over during evolution's time, providing interesting observational outcomes \cite{vie}.
Apart from this, understanding low mass stars' evolution means that we understand more about galaxies, as about $70\%$ of their stars are those particular ones.

In what follows, we will start with the introduction of non-relativistic stars in a general class of ST theories in section \ref{sec2}. In section \ref{sec3} we will analyse the PMS phase - we will mainly focus on the gravitational contraction and lithium burning in DHOST theories, as well as on the Schwarzschild criterion widely used in the stellar modelling. In the last section we draft our conclusions. We will also discuss possible tests of gravity with the use of fully convective stars.

\section{Non-relativistic stars in ST gravity}\label{sec2}

\subsection{Hydrostatic equilibrium equation for non-relativistic stars}
{In this section, we will discuss the equation of state (EoS) for non-relativistic stars in Horndeski Gravity. Let us begin by briefly introducing the theory. The Horndeski theory of gravity is the most general ST theory containing up to second order field equations and it is described by the following Lagrangian \cite{galileon2} }
{
\begin{align} 
{\cal L}&=G_2(\phi,X)-G_3(\phi,X)\Box\phi + G_4(\phi,X)R \nonumber \\
&+G_{4X}  \left[(\Box\phi)^2-\phi^{\mu\nu}\phi_{\mu\nu}\right]
 +G_5(\phi,X) G^{\mu\nu}\phi_{\mu\nu} \nonumber
 \\ & -\frac{G_{5X}}{6}\left[
  (\Box\phi)^3-3\Box\phi\phi^{\mu\nu}\phi_{\mu\nu}
  +2\phi_{\mu\nu}\phi^{\nu\lambda}\phi_\lambda^\mu
  \right], \label{Horndeski}
\end{align}
where $R$ is the Ricci tensor, $G_{\mu\nu}$ is the Einstein tensor,
$\phi_\mu:=\nabla_\mu\phi$, $\phi_{\mu\nu}:=\nabla_\mu\nabla_\nu\phi$, $X:=-g^{\mu\nu}\phi_\mu\phi_\nu/2$, $f_X:=\partial f/\partial X$, $f_\phi:=\partial f/\partial \phi$ and $G_2$, $G_3$, $G_4$, and $G_5$ are arbitrary functions of the fields $\phi$ and $X$.
Any second-order ST theory can be reproduced from \eqref{Horndeski} via a suitable choice of functions $G_i$. Therefore, we can say that Horndeski theory encloses all second-order ST theories.\\
The advantage of working with second-order ST theories lies in the fact that those theories do not propagate ghostly degrees of freedom. However, it is possible to have healthy higher-derivative theories when the system is degenerate because, in such cases, we can eliminate the higher derivatives in the equations of motion \cite{gley,rham,crem}. The set of theories that falls in this classification is referred as DHOST theories. This suggests that we can work with theories beyond Horndeski, i.e., DHOST theories that extend Horndeski gravity and pass the test provided by the GW170817 event \cite{ligo,ligo2} (therefore, we focus on particular sub-classes of DHOST theories). Moreover, the only DHOST theories that do not suffer from instabilities are those that can be connected to Horndeski gravity by a disformal invertible transformation \cite{ben1} (see \cite{Kobayashi} for a review).}

Let us now briefly recall the basic equations describing a non-rotating star in a hydrostatic equilibrium used in the further part. Since we are interested in low-mass stars, that is, stellar objects with masses not exceeding $0.6M_\odot$, their convective interior is well-modelled by the polytropic equation of state with $n=3/2$
\begin{equation}\label{pol}
p=K\rho^\frac{n+1}{n}
\end{equation}
where $n$ is called the polytropic index while the parameter $K$ in the simplest model is a constant. However, as seen in the next section, $K$ can also include an information about the gas mixture of the stellar material, and electron degeneracy - being very important when one already deals with very low-mass stars.

{Before going further, we should discuss the validity of the polytropic and ideal gas EoS's in the framework of modified gravity. Although it was demonstrated (see e.g. section III.3 in \cite{awmicro}) that one should take into account the effects of (modified) gravity in Fermi EoS\footnote{Fermi EoS describes particles with the Fermi-Dirac statistics which plays a crucial role in the contracting objects such as pre-Main Sequence stars, brown dwarfs and giant gaseous planets.}, in the case of low temperatures and non-relativistic mass and hydrostatic equilibrium equations those effects are insignificant. In the mentioned paper, the derivations are general and can be applied to any theory of gravity. However, if one considers relativistic stars, polytrope should not be used (instead, one should use, in the simplest case, Chandrasekhar EoS). In our work we consider the non-relativistic limit of beyond - Horndesky theory, that is, non-relativistic hydrostatic equilibrium equations, therefore the mentioned result also applies to this particular theory of gravity.}

{
 Let us however notice that usually one thinks about the very simple polytropic EoS, when $K$ is a constant. As mentioned above, one can indeed "hide" many interesting effects in it, such as e.g. the electron degeneracy, crucial in our analysis of contracting stars, strongly coupled plasma, finite gas temperatures with phase transition points between metallic hydrogen and molecular state \cite{burrows,auddy}, and finite strains \cite{seager}, to mention just a few of them. Moreover, a mixture of different polytropic EoS's with different polytropic indices $n$ (and also ideal gas) can be rewritten as polytropic EoS with $n=3/2$ (see e.g. \cite{anetaJup,auddy,pachol}).
 Regarding the polytropic index $n$, it is a well-known fact that fully convective stars are described by a polytropic EoS with $n=3/2$ if those stars are not more massive that $\approx 0.35M_\odot$ \cite{chabrier} and because the polytropic form can be safely used in modified gravity, we should also have this mass limit in mind. Indeed, in the further part of this work, we will be interested in stars below that threshold.}

{
 On the other hand, the atmosphere of the stellar objects which we will study in this paper is modelled by the ideal gas (see eq. \eqref{ideal}), which in the framework of modified or quantum gravity does not acquire any modifications \cite{moussa,ali}.
 }

{
The ideal gas is the simplest approximation used to describe the matter behaviour in the stellar atmosphere. Even if we take into account ionization of hydrogen and helium with phase transition points between metallic hydrogen and molecular state, in the case when degeneracy does not play so important role as it happens in the atmosphere of the considered objects, the equation of state reduces to the ideal gas form \cite{auddy}.}

{We can now turn our attention to the modifications of Horndeski theory to the hydrostatic equilibrium equation and mass function. Taking the Lagrangian \eqref{Horndeski} into consideration, it can be shown that the hydrostatic equilibrium equation is modified as follows \cite{broken}
\begin{equation}\label{hydroeq}
\frac{\mathrm{d} p}{\mathrm{~d} r}=-\frac{G_{\mathrm{N}} M(r) \rho(r)}{r^{2}}-\frac{\Upsilon}{4} G_{\mathrm{N}} \rho(r) M^{\prime \prime}(r)
\end{equation}
while the mass function, given by
\begin{equation} \label{masseq}
    \frac{dM}{dr}=4\pi r^2 \rho(r),
\end{equation}
is unaffected by the theory.}

The above equation of state \eqref{pol}, together with the modified hydrostatic equilibrium equation \eqref{hydroeq}
and the mass function \eqref{masseq}
provide that the modified Lane-Emden equation (LEE) for Horndeski gravity is \cite{broken,sak1,sak2}
\begin{equation} \label{LE}
    \frac{1}{\xi^{2}} \frac{\mathrm{d}}{\mathrm{d} \xi}\left[\left(1+\frac{n}{4} \Upsilon \xi^{2} \theta^{n-1}\right) \xi^{2} \frac{\mathrm{d} \theta}{\mathrm{d} \xi}+\frac{\Upsilon}{2} \xi^{3} \theta^{n}\right]=-\theta^{n}.
\end{equation}
{The modified LEE can be obtained by taking the hydrostatic equilibrium equation \eqref{hydroeq} into consideration and writing the radius as $r=r_c \xi$, with $r_c = (n+1) P_c/4 \pi G_N \rho_c^2$. The pressure and density are rewritten in terms the central pressure $ P_c $ and density $\rho_c $ (which are related by the polytropic equation of state \eqref{pol}) as $P= P_c \theta^{n+1}(\xi)$ and $\rho=\rho_c \theta^n(\xi)$, respectively.
}

The solutions of the modified Lane-Emden equation (\ref{LE}) provide
the star's mass, radius, central density, and temperature via the well-known expressions (see e.g \cite{weinberg})
\begin{align}
 M&=4\pi r_c^3\rho_c\omega_n,\\
 R&=\gamma_n\left(\frac{K}{G}\right)^\frac{n}{3-n}M^\frac{n-1}{n-3} \label{radiuss},\\
 \rho_c&=\delta_n\left(\frac{3M}{4\pi R^3}\right) \label{rho0s} ,\\
 T_c&=\frac{K\mu}{k_B}\rho_c^\frac{1}{n}\theta_n \label{temps},
\end{align}
where $k_B$ denotes the Boltzmann constant and $\mu$ the mean molecular weight.
The constants $\omega_n$, $\gamma_n$ and $\delta_n$ are defined as:
\begin{equation}
    \omega_n  = -\xi_R^2 \frac{d\theta}{d\xi} \Big|_{\xi=\xi_R},
\end{equation}
\begin{equation}
    \gamma_n = (4\pi)^{\frac{1}{n-3}}(n+1)^{\frac{n}{n-3}}\omega_n^{\frac{n-1}{n-3}} \xi_R.
\end{equation}
\begin{equation}
    \delta_n = -\xi_R \left(3 \frac{d\theta}{d\xi} \Big|_{\xi=\xi_R} \right)^{-1},
\end{equation}
Let us mention that the parameter $\Upsilon$ has been already constrained with the use of data related to different astrophysical probes \cite{chow,rosyadi}. In what follows, we will focus on its values from the range $-\frac{2}{3}<\Upsilon\lesssim 0.3$ given by \cite{sak1,rosyadi}, where a similar class of objects were considered. More restrict constraints are given in \cite{saltas}, of the order of magnitude $10^{-4}$, which were obtained by studying seismic properties in the Sun, which is modelled in different way (there are more layers) that one models low-mass stars. Since the effects such as rotation, magnetic field and evolution of the electron degeneracy are not taken into account there, those bounds are not definitive. Secondly, the provided bounds (at $2\sigma$) were obtained with studying only one object. However, a large statistics and the current data can improve the confidence. Because of that fact, and a similarity to the objects used to constrain DHOST theory, we will also consider larger bounds (at $5\sigma$) given in \cite{rosyadi}, for which the effects of the scalar field are more evident.

\section{Pre-main sequence phase}\label{sec3}
In what follows, we will consider two processes related to the PMS phase of the stellar evolution. Roughly speaking, a PMS star contracts until one of the three processes happens in its core: radiative core develops (a special case related to it is discussed in the subsection \ref{radcore}), hydrogen starts being burnt, or electron degeneracy pressure is already high enough to stop the gravitational contraction. Before any of the mentioned processes happens, the baby star follows the Hayashi track \cite{hayashi}, that is, an evolutionary path placed in the cold region of the Hertzsprung-Russell (HR) diagram, given by a curve being almost perpendicular to the MS. The PMS stars on Hayashi stars are fully convective (apart from radiative envelopes), therefore their interiors are well described by the polytropic equation of state.

Depending on its mass, the star can start developing radiative core because of growing luminosity and/or opacity, or hydrogen ignition starts, as the core's conditions are sufficient for it. In the first case the star's effective temperature grows at almost constant luminosity --- the PMS stars enter the much shorter phase and follow the Henyey tracks \cite{hen1,hen2,hen3}. This evolutionary scenario happens for stars with masses bigger than $\sim 0.6M_\odot$ for Newtonian gravity; stars with lower masses reach the MS being still fully convective. The star, independently if it follows Hayashi or Henyey track, which have a core hot enough to start burning hydrogen, moves on to the MS phase. In the case when a fully convective star does not have sufficient conditions to ignite hydrogen, such an object will further contract and increase the electron degeneracy pressure, which will finally stop the gravitational contraction. Since there is no relevant energy production, such aborted stars, called brown dwarfs, will cool down with time. Cooling processes of brown dwarfs and giant planets have been studied in \cite{kart} in Horndeski gravity, while the onset of hydrogen burning in low-mass stars have been studied in \cite{sak1,sak2}.

In the following work, we are mainly focused on three processes related to the stellar evolution: Hayashi tracks, that is, the ongoing contraction, lithium burning during that phase, and a case of a maximal mass of a MS star. Such a star will
be modelled as a ball made of fully ionized monatomic gas with mean molecular weight $\mu$, surrounded by a radiative atmosphere. Therefore, we will deal with two temperatures: the one of the interior, denoted by $T$, and the effective one $T_\text{eff}$, which we assume to be the temperature of the photosphere --- a region of the atmosphere placed approximately at $r\approx R$, where $R$ is the radius of the star. In the atmosphere, one deals with radiative processes; the most important and difficult part of atmosphere modelling is related to the absorption process. In order to be able to carry our studies on a theoretical level, and to focus only on modified gravity effects, we will use a Kramer law, which is a simple relation between the opacity, pressure $p$, and temperature $T$: 
\begin{equation}\label{abs}
 \kappa_\text{abs}= \kappa_0 p^w T^v,
\end{equation}
where $\kappa_0$, $w$, and $v$ are constants whose values depend on the atmosphere composition and temperature range.

\subsection{Contracting to the Main Sequence}
For our toy model star modelled as mentioned above, the polytropic equation of state (\ref{pol}), with the use of the ideal gas relation ($N_A$ and $k_B$ are the Avogadro and Boltzmann constants, respectively):
\begin{equation}\label{ideal}
 \rho=\frac{\mu p}{N_A k_B T},
\end{equation}
can be rewritten in a more suitable form for the further purposes
\begin{equation}\label{eos}
 p=\left(\frac{N_Ak_B}{\mu}\right)^{1+n}\frac{T^{1+n}}{K^{n}},
\end{equation}
where $K$ is given by the solutions of the modified Lane-Emden equation (\ref{LE})
\begin{equation}\label{ka}
 K=\left[\frac{4\pi}{\xi_R^{n+1}(-\theta'_n(\xi_R))^{n-1}}\right]^\frac{1}{n}\frac{G_N}{n+1}M^{1-\frac{1}{n}}R^{\frac{3}{n}-1}.
\end{equation}
On the other hand, we also need expressions which will allow us to describe the photopshere's and atmosphere's characteristics. The photosphere can be defined as a surface with temperature $T_\text{eff}$ for which the optical depth $\tau$ takes the value $2/3$:
\begin{equation} \label{eq:od}
 \tau(r)=\kappa\int_r^\infty \rho dr=\frac{2}{3}.
\end{equation}
Moreover, the photopshere quantities satisfy the Stefan-Boltzmann law (let us recall that photosphere is assumed to lie at $r\approx R$), since it is a visible surface from which the radiation is emitted into space ($\sigma$ is the Stefan-Boltzmann constant):
\begin{equation}\label{stef}
    L=4\pi\sigma R^2T^4_\text{eff}.
\end{equation}
The atmosphere instead, as already mentioned, is mainly described by the opacity; in our model we will consider a simple power-law form (\ref{abs}). The baby stars following Hayashi tracks can be found in the right hand side region of the HR diagram - that is, they are 
cool, gaseous objects with the surface temperatures being in the range $3000\lesssim T\lesssim6000$K, such that
its surface layer
is dominated by H$^{-}$ opacity \cite{hansen}. With hydrogen mass fraction $X\approx0.7$, the H$^{-}$ opacity is given by
\begin{equation}\label{hydro}
 \kappa_{H^-}=\kappa_0 \rho^\frac{1}{2}\,T^9\,\,\text{cm}^2\text{g}^{-1},
\end{equation}
where $\kappa_0\approx2.5\times10^{-31} \left(\frac{Z}{0.02}\right)$.
Then, the usual metal mass fraction $Z$ lies in the range $0.001\lesssim Z\lesssim0.03$. The solar metallicity is $Z=0.02$. 
Before going further, let us notice that in the case of the ideal gas, the opacity (\ref{hydro}) can be expressed as 
\begin{equation}\label{hydro2}
 \kappa_{H^-}=\kappa_g p^\frac{1}{2}\, T^{8.5}\,\,\text{cm}^2\text{g}^{-1},
\end{equation}
where $\kappa_g=\kappa_0\left(\frac{\mu}{N_Ak_B}\right)^\frac{1}{2}\approx1.371\times10^{-33}Z\mu^\frac{1}{2}$. 

Assuming that the surface gravity is constant,
\begin{equation}\label{assum}
     g=\frac{G_NM(r)}{r^2}=\text{const}
\end{equation}
we can rewrite the hydrostatic equilibrium equation as
\begin{equation}\label{surface}
    \frac{dp}{dr}=-\frac{G_NM(r)}{r^2}\rho(r)\left(1+\frac{\Upsilon}{2}\right),
\end{equation}
which can be integrated with $r=R$ and $M=M(R)$ when applied to (\ref{eq:od}). {Let us emphasize here that the above assumption is valid as long as we work on the non-relativistic limit of the theory, as the only modification to the EoS, in this case, will be parametrized by $\Upsilon$ \cite{broken}. In fact, in the relativistic limit, $G$ will be dynamical and the approximation \eqref{assum} cannot be used.}
With the use of the absorption law (\ref{hydro2}), the photopsheric pressure has the following form
\begin{equation}\label{fotos}
 p_\text{ph}=8.12\times10^{14}\left(\frac{ M\left(1+\frac{\Upsilon}{2}\right)}{L T_\text{ph}^{4.5}Z\mu^\frac{1}{2}}\right)^\frac{2}{3},
\end{equation}
where the Stefan-Boltzmann law (\ref{stef}) after identifying that $ T_{\text{eff}{\mid_{r=R}}}\equiv T_\text{ph}$ was adopted.

Let us come back to the equation (\ref{eos}). Considering the fully convective case, that is, $n=3/2$, and taking it on the photosphere, we find that (we will skip the index $N$ in the further part in $G_N$)
{\small
\begin{equation}
 T_{\text{eff}{\mid_{r=R}}}=\left(\frac{\mu}{N_Ak_B}\right)^{-\frac{2}{3}}\left(\frac{4\pi}{\xi (-\theta')^\frac{1}{2}}\right)^\frac{2}{5}
 \left(\frac{2G}{5}\right)^\frac{3}{5}M^\frac{1}{5}R^\frac{3}{5}p_\text{ph}^\frac{2}{5}.
\end{equation}
}
To get rid of the radius $R$ in the above equation, let us use again the Stefan-Boltzmann law; after inserting numerical values of the constants, the photopsheric temperature can be expressed as function of the luminosity, mass, and photospheric pressure:
\begin{equation}
  T_\text{ph}=9.196\times10^{-6}\left( \frac{L^\frac{3}{2}Mp_\text{ph}^2\mu^5}{-\theta'\xi_R^5} \right)^\frac{1}{11}.
\end{equation}
The photospheric pressure must be identified with the gravitational pressure taken on the photosphere given by the derived equation (\ref{fotos}). Using it in the above equation, we can finally write the expression for the Hayashi track:
{\small
\begin{equation}\label{hay}
 T_\text{ph}=2487.77\mu^\frac{13}{51}\left( \frac{L}{L_\odot}  \right)^{\frac{1}{102}}\left( \frac{M}{M_\odot}  \right)^{\frac{7}{51}}
 \left( \frac{\left(\frac{\left(1+\frac{\Upsilon}{2}\right)}{Z}\right)^\frac{4}{3}}{\xi_R^5\sqrt{-\theta'}} \right)^\frac{1}{17}\text{K},
\end{equation}
}
 where $L_\odot$ and $M_\odot$ are the solar luminosity and mass, respectively.
 
\begin{figure}[t]
\centering
\includegraphics[scale=.55]{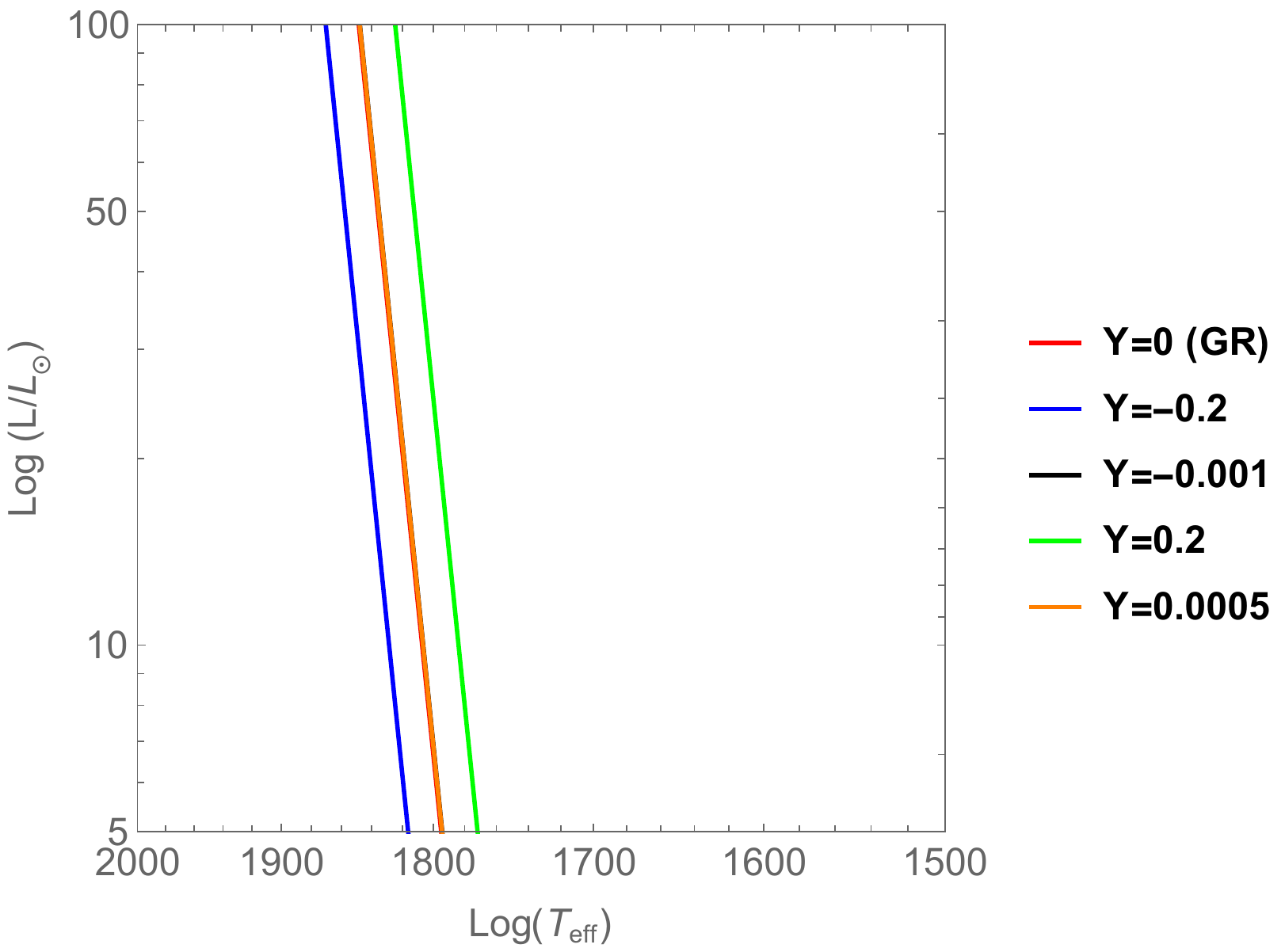}
\caption{[color online] The Hayashi tracks of a star with mass $M=0.25M_\odot$, metallicity $Z=0.02$, and chemical composition $\mu=0.618$ with respect to
a few values of the parameter $\Upsilon$, given by the equation (\ref{hay}).}
\label{fig}
\end{figure}

 For a given star's mass $M$, mean molecular weight $\mu$, and metallicity $Z$, the above equation provides an evolutionary track of the PMS star. Although our derivation suffers a number of assumptions which allowed us to simplify the equations to work it out analytically, the obtained result clearly demonstrates the dependence on an applied model of gravity. A few curves corresponding to different values of the parameter ($\Upsilon=0$ gives Newtonian gravity curve) for a star with mass $M=0.25M_\odot$, mean molecular weight $\mu=0.618$, and solar metallicity $Z=0.02$ are given in the figure \ref{fig}. Unfortunately, our simple relation (\ref{hay}) does not reflect the importance of the metallicity \cite{met} - in the considered toy-model, the curves are only slightly shifted for different values instead of changing the curve's shape, as it happens in more realistic models, when one properly treats the atmosphere's opacity problem.

\subsection{Lithium burning}

The lithium burning process, which occurs at the center of the star, induces a flux of lithium-rich fluid to the center of the star and lithium-poor fluid to its outer regions. This process is possible as long as the mixing timescale is much smaller than the contraction and lithium destruction times. Moreover, it is responsible for maintaining  the depletion rate constant throughout the star as the total quantity of lithium reduces over time. Proton-capture reactions also play a role in the process and should be taken into account. For a star with mass $M$ and hydrogen fraction $X$ we can write the depletion rate as
\begin{equation} \label{reac}
 M\frac{\text{d}f}{\text{d}t}=-\frac{Xf}{m_H}\int^M_0\rho\langle\sigma v\rangle dM,
\end{equation}
where the non-resonant reaction rate depends on the temperature range; for the range of temperature appropriate to our analysis, $T<6\times 10^6\text{K}$, it is given by
\begin{equation} \label{reaction}
 N_A\langle\sigma v\rangle=Sf_\text{scr} T^{-2/3}_{6}\text{exp}\left[-aT_{6}^{-\frac{1}{3}}\right]\;\frac{\text{cm}^3}{\text{s g}},
\end{equation}
with $T_{6}\equiv T/10^6\text{K}$, $f_\text{scr}$ being the screening correction factor and $S$ and $a$ being dimensionless parameters related to the proton-capture rate. Considering our temperature range and the reaction $^7\text{Li}(p,\alpha)\,^4\text{He}$, the proton-capture rate parameters are given by $S=7.2\times10^{10}$ and $a=84.72$ \cite{usho,cf,raimann}.

We are concerned about low-mass stars, which can be theoretically described by a polytropic equation of state for $n=3/2$ since they are fully convective. Therefore, the temperature and the density are expressed as $T=T_c\theta(\xi)$ and 
$\rho=\rho_c\theta^{3/2}(\xi)$, respectively. Due to the modification in the LEE \eqref{LE}, $\delta$, $\xi_R$ and $\theta'$ will be modified accordingly. Consequently, the central temperature $T_c$ and central density $\rho_c$ will be also modified:
{\small
\begin{align}\label{ctemp}
 T_c=&1.15\times 10^6 \left(\frac{\mu_\text{eff}}{0.6}\right)\left(\frac{M}{0.1M_\odot}\right) \left(\frac{R_\odot}{R}\right)
 \frac{\delta^\frac{2}{3}}{\xi_R^\frac{5}{3}(-\theta'(\xi_R))^\frac{1}{3}}\text{K}\\
 \rho_c=&0.141\left(\frac{M}{0.1M_\odot}\right) \left(\frac{R_\odot}{R}\right)^3\delta\,\frac{\text{g}}{\text{cm}^3}.
\end{align}
}
If we take into account an arbitrary degeneracy degree $\eta$ and a mean molecular weight $\mu_\text{eff}$, we find that the radius is given by
\begin{equation}
 \frac{R}{R_\odot}\approx\frac{7.1\times10^{-2}\gamma}{\mu_\text{eff}\mu_e^\frac{2}{3}F^\frac{2}{3}_{1/2}(\eta)}
 \left(\frac{0.1M_\odot}{M}\right)^\frac{1}{3}\label{Rpol},
\end{equation}
where $F_n(\eta)$ is the $n$th order Fermi-Dirac function.  

Changing to spatial variables in \eqref{reac} and using the reaction rate \eqref{reaction} together with the polytropic equation of state for the energy density, we obtain
\begin{align}\label{rate}
  \frac{\text{d}}{\text{d}t}\text{ln}f&=-\frac{4\pi X}{\xi_R^3 }\frac{\rho^2_c R^3}{M}\frac{S}{N_A m_H}\left(\frac{u}{a}\right)^2 \nonumber\\
  &\times\int_0^{\xi_R}f_\text{scr} \xi^2\theta^\frac{7}{3} \text{exp}(-u\theta^{-1/3})d\xi\,\,\,\,  \frac{1}{\text{s}},
\end{align}
where we defined $u\equiv aT_{c6}^{-1/3}$ for convenience.
To proceed further, we will need the solutions of the equation \eqref{LE}. In general, it is not possible to obtain exact solutions to the LEE for the considered value of the polytropic parameter $n$. Therefore, one has to take the approximate near center solution. Such approximation is justified by the fact that the burning process occurs at the central region of the star. Some theories with modified LEE have the same solutions as the original one \cite{aneta3}. For scalar-tensor theories, however, the approximate solution to \eqref{LE} for $n=3/2$ is given by \cite{sak1,sak2}
\begin{equation}\label{solution}
    \theta( \xi \approx 0) \approx 1 - \left( 1+ \frac{3 \Upsilon}{2}  \right)\frac{\xi^2}{6} \approx \exp\left[  - \left( 1+ \frac{3 \Upsilon}{2}  \right)\frac{\xi^2}{6} \right],
\end{equation}
where the boundary conditions $\theta(0)=1$ and $\theta'(0)=0$ were used. In this case, we can clearly see that the solution \eqref{solution} depends on the theory parameter $\Upsilon$. We expect that this dependency will be present in the lithium depletion rate as well. Using the solution \eqref{solution} and applying the numerical constants to (\ref{rate}), we obtain
\begin{align}
   & \frac{d \ln f}{dt} =- 6.54 S f_{scr} \xi_R^2  \left(\frac{X}{0.7}\right)\left(\frac{0.1 M_\odot}{M} \right)^2 \left(\frac{0.6}{\mu_{eff}} \right)^3 \nonumber \\
    &\times (-\theta'(\xi_R))  a^7 u^{-17/2} e^{-u}\left(1+ \frac{7}{u}\right)^{-3/2}  
    \left(1 + \frac{3 \Upsilon}{2} \right)^{-3/2},
\end{align}
which confirms that the depletion rate depends on the theory parameter $\Upsilon$.
From the Stefan-Boltzmann equation and the virial theorem we can show that the star's luminosity obeys the relation
\begin{equation}\label{stefan}
    L= 4 \pi R^2 \sigma T^4_{eff} = -\frac{3}{7} \frac{GM^2}{R^2}\frac{dR}{dt},
\end{equation}
from which we can get the radius and luminosity as functions of time:
{\small
\begin{equation}
    \frac{R}{R_\odot} = 0.85 \left(\frac{M}{0.1M_\odot} \right)^{\frac{2}{3}} \left(\frac{3000K}{T_{eff}}\right)^{\frac{4}{3}} \left(\frac{Myr}{t} \right)^{\frac{1}{3}}, 
\end{equation}
\begin{equation}
    \frac{L}{L_\odot} = 5.25 \times 10^{-2}  \left(\frac{M}{0.1M_\odot} \right)^{\frac{4}{3}} \left(\frac{T_{eff}}{3000K} \right)^{\frac{4}{3}} \left(\frac{Myr}{t} \right)^{\frac{2}{3}}.
\end{equation}
}

The contraction time is given in terms of the central temperature $T_c$:
\begin{eqnarray}\label{cont}
    t_{cont} &=& -\frac{R}{dR/dt} \nonumber \approx 31.17 \left(\frac{3000K}{T_{eff}}\right)^4 \left(\frac{0.1 M_\odot}{M} \right)  \left(\frac{0.6}{\mu_{eff}} \right)^3 \\
   &\times&  \left( \frac{T_c}{3 \times 10^6K}\right)^3 \frac{\xi^5_R(-\theta'(\xi_R))}{\delta^2} Myr. 
\end{eqnarray}
We can now rewrite the depletion rate as an integral in $u$. For this, we need to notice from \eqref{ctemp} and from the definition $u = aT_{c6}^{-1/3}$ that\footnote{Let us comment that from now we are interested in stars with masses $>0.2M_\odot$ such that the change in the electron degeneracy in negligible with respect to the change in the star's size.}
\begin{equation}
    \frac{d}{dt} \ln f \approx \frac{d \ln f}{du} \frac{\partial u}{\partial R} \dot{R} = \frac{d \ln f}{du} \frac{u \dot{R}}{3R}.
\end{equation}
Therefore, the depletion rate can be rewritten as
{\small
\begin{eqnarray}
      \frac{d \ln f}{du} &=&  5.6 \times 10^{14} T_{eff}^{-4}  \left(\frac{X}{0.7}\right)\left(\frac{0.1 M_\odot}{M} \right)^3 \left(\frac{0.6}{\mu_{eff}} \right)^6 \\
      &\times& S f_{scr}  a^{16} u^{-37/2} e^{-u}\left(1- \frac{21}{2u}\right) \nonumber \\
      &\times&
      \left(1 + \frac{3 \Upsilon}{2} \right)^{-3/2} \frac{\xi_R^7(-\theta'(\xi_R))^2}{\delta^2}. \nonumber
\end{eqnarray}
}

We can now obtain the depletion $\mathcal{F}$ as a function of $u$, that is, as a function of the central temperature $T_{c}$. For this purpose, we just need to integrate from $u_0=+\infty$ to $u$, with the initial abundance being given by $f_0$. Thus,
\begin{eqnarray}\label{ratio}
       \mathcal{F} &=& \ln \frac{f_0}{f}  =  5.6 \times 10^{14} T_{eff}^{-4}  \left(\frac{X}{0.7}\right)\left(\frac{0.1 M_\odot}{M} \right)^3 \left(\frac{0.6}{\mu_{eff}} \right)^6 \nonumber\\ 
       &\times& S f_{scr}  a^{16}g(u) \left(1 + \frac{3 \Upsilon}{2} \right)^{-3/2} \frac{\xi_R^7(-\theta'(\xi_R))^2}{\delta^2},
\end{eqnarray}
where we defined $g(u)= u^{-37/2}e^{-u}-29\Gamma(-37/2,u)$, with $\Gamma (-37/2,u)$ being the upper incomplete gamma function. 
Following the same procedure, it is possible to obtain the depletion rate for resonant rates \cite{usho} 
\begin{eqnarray} \label{ratio2}
       \mathcal{F} &=& \ln \frac{f_0}{f} = 5.6 \times 10^{14} T_{eff}^{-4}  \left(\frac{X}{0.7}\right)\left(\frac{0.1 M_\odot}{M} \right)^3 \left(\frac{0.6}{\mu_{eff}} \right)^6 \nonumber\\ 
       &\times& S f_{scr}  a^{18-3j}\Bar{g}(u) \left(1 + \frac{3 \Upsilon}{2} \right)^{-3/2} \frac{\xi_R^7(-\theta'(\xi_R))^2}{\delta^2}, 
\end{eqnarray}
where $j=2/3$ corresponds to a non-resonant reaction and 
$\Bar{g}(u) = u^{-41/2+3j}e^{-u} - \frac{(68-15j)}{2} \Gamma \left(-\frac{41}{2} +3j,u\right)$. Therefore, it is possible to have a relation $u(\mathcal{F})$ which allows us to determine the central temperature $T_c$  for a given depletion $\mathcal{F}$. Such relation could be solved numerically and fitted to data, but the approximated value of central temperature can also be obtained from the time of depletion.

 {For a star described by a polytropic equation of state with $n=3/2$, the contraction time $t_{cont}$ is comparable to the destruction time $t_{dest}$ for which lithium is depleted if we deal with a mild degeneracy:}
\begin{eqnarray}\label{destr}
       t_{dest} &=& \frac{m_P}{X \rho <\sigma v>} = 4.92 \times 10^{-7} \left(\frac{M}{0.1 M_{\odot}}\right)^2 \left(\frac{\mu_{eff}}{0.6} \right)^3 \nonumber \\
       &\times& \frac{T_{c6}^{-\frac{7}{3}}}{S f_{scr}} e^{\frac{a}{T_{c6}^{1/3}}} \frac{\delta}{\xi_R^5 (-\theta'(\xi_R))} yr. 
\end{eqnarray}
 {
It was demonstrated \cite{usho} that in case when the star is described by the polytrope with $n=3/2$ the equality of those timescale are indeed comparable if degeneracy can be neglected, that is, when the change in the electron
degeneracy in negligible with respect to the change in the star's radius, $\dot\mu_{eff}<\dot R$. 
In the case when $n=3/2$ does not hold anymore, which happens when the core starts being radiative, those timescales are not comparable. Since we limited our calculations to stars with masses beyond this threshold, we can take the approximation $t_{cont}=t_{dest}$ to obtain the following relation for the central temperature}
\begin{eqnarray}\label{central}
        \frac{a}{T_{c6}^{1/3}} &=& 28.48 + \ln (S f_{scr}) - 4\ln\left(\frac{T_{eff}}{3000K}\right) \nonumber \\ 
        &-&3 \ln \left(\frac{M}{0.1M_{\odot}} \right) +\frac{16}{3} \ln T_{c6} -6\ln \left(\frac{\mu_{eff}}{0.6}\right)\nonumber \\
    &+&  \ln \left(\frac{\xi_R^{10}(-\theta'(\xi_R))^2}{\delta^3} \right),
\end{eqnarray}
from which we have also obtained, given in the Table \ref{tab_age}, the age, radius, and luminosity of a $0.35M_\odot$ star for various values of $\Upsilon$.

 {
On the other hand, in the case when degeneracy cannot be neglected, the contraction time is longer than the destruction one. Let us notice that $t_\text{dest}$ strongly depends on the central temperature. It is so because the reaction rate is very sensitive to even slight changes in the central temperature, at the same time being insensitive to small uncertainties in the constitutive physics. For $T_c$ high enough the element can be depleted before reaching the Main Sequence, so then $t_\text{dest}\neq t_\text{cont}$. As demonstrated \cite{usho}, although contraction (shrinking radius) increases the central temperature at the beginning, in the case when degeneracy becomes significant, it has a non-trivial effect on the central temperature (it decreases $T_c$). Because of that fact, we limit our consideration to the stars with masses $M>0.02M_\odot$.
}

\begin{table}[p]
\begin{center}
\begin{tabular}{||c | c | c | c | c||}
 \hline
 $\Upsilon$ & $T_c/10^6K$ & $t(Myr)$ & $R/R_{\odot}$ & $L/L_{\odot}$ \\ [0.5ex] 
 \hline\hline
 -0.6 & 2.75 & 51.65 & $0.245$ & $0.175$ \\
-0.3 & 2.86 & 19.10 & $0.242$ & $0.170$ \\
\hline
0 (GR) & 2.97 & 17.21 & 0.239 & $0.166$ \\
\hline
0.3 & 2.98 & 11.13 & $0.239$ & $0.166$ \\
0.6 & 3.03 & 7.95 & $ 0.238 $& $0.164$
 \\ [1ex] 
 \hline
\end{tabular}
\caption{Numerical values for the central temperature (in $10^6$K), age (in Myr), radius (with respect to $R_{\odot}$) and luminosity (with respect to $L_{\odot}$) for different values of the parameter $\Upsilon$. The following values were used for the star’s mass, effective temperature,
hydrogen mass fraction, and mean molecular weight: $M=0.35M_{\odot}$, $T_{eff} = 3500K$, $X=0.7$ and $\mu_{eff}=0.6$.}\label{tab_age}
\end{center}
\end{table}

\subsection{Reaching the Main Sequence as a fully convective star}\label{radcore}

In our assumption we treat the star as a fully convective ball surrounded by a radiative atmosphere. Since we expect that the heat transfer changes from the convective process to the radiative one in the neighbourhood of the surface, we may write the hydrostatic equilibrium equation there as \eqref{surface}
\begin{equation}
    \frac{dp}{dr}=-\frac{G_NM(r)}{r^2}\rho(r)\left(1+\frac{\Upsilon}{2}\right).
\end{equation}
The heat transport with respect to radiative/conductive process is given by
\begin{equation}\label{gradM}
    \frac{\partial T}{\partial M}=-\frac{3}{64\pi^2 b c}\frac{\kappa_\text{rc}L}{r^4 T^3}
\end{equation}
which can be rewritten with the use of (\ref{surface}) as
\begin{equation}\label{gradT}
    \frac{\partial T}{\partial p}= \frac{3\kappa_\text{rc}L}{16\pi G M b c T^3}\left(1+\frac{\Upsilon}{2}\right)^{-1},
\end{equation}
such that the temperature gradient 
\begin{equation*}
    \nabla_\text{rad}:= \left( \frac{d\textrm{ln} T}{d\textrm{ln} p} \right)_\text{rad}
\end{equation*}
is given by\footnote{Let us notice that this relation works only for fully convective stars with a radiative envelope, for which the boundary lies at $r\approx R$. In the case of higher masses, when one deals with a more complicated internal structure, one should use the equation \eqref{hydroeq} in \eqref{gradM}.}
\begin{equation}\label{schw}
    \nabla_\text{rad}=\frac{3\kappa_\text{rc}Lp}{16\pi G M b c T^4}\left(1+\frac{\Upsilon}{2}\right)^{-1}.
\end{equation}

Using the Kramers law (\ref{abs}) for the total bound-free and free-free opacities \cite{hansen} for which $w=1$ and $v=-4.5$, while $\kappa_0$'s are given by
\begin{align}
 \kappa_0^\text{bf}&\approx 4\times10^{25}\mu \frac{ Z(1+X)}{N_Ak_B}\text{cm}^2\text{g}^{-1},\label{bf}\\
  \kappa_0^\text{ff}&\approx 4\times10^{22}\mu\frac{(X+Y)(1+X)}{N_Ak_B}\text{cm}^2\text{g}^{-1},\label{ff}
\end{align}
respectively, and together with the equation \eqref{eos} one can write
\begin{eqnarray}
     \nabla_\text{rad}=5.9\times 10^{-3}  \frac{N_A^5 k_B^5 \kappa_0 L}{b c G^4 \mu ^5M^2 R^3 T^{3.5}}  \frac{\xi_R^5(-\theta'(\xi_R))}{\left(1+\frac{\Upsilon}{2}\right)} .
\end{eqnarray}
The homology contraction argument and the equation \eqref{temps} for the central temperature allow us to write
\begin{equation}
    T_c = 6.679 \times 10^{-16} \frac{\mu \delta_{3/2}^{2/3}}{\xi_R^{5/2} (-\theta'(\xi_R))^{1/3}} \frac{M}{R},
\end{equation}
which we apply, together with the Boltzmann law, into the gradient $\nabla_{rad}$:
\begin{eqnarray} \label{nabla}
\nabla_{rad}&=&
2.79\times 10^{-148}  \frac{k_B^{8.5} N_A^{8.5}\xi_R^{10.83}(-\theta'(\xi_R))^{2.167} \kappa_o L ^{1.25}}{b c G^{7.5} \delta_{3/2}^{2.33} \mu^{8.5}M_{-1}^{5.5}T_{\text{eff}}} \nonumber \\ 
&\times& \left(1+\frac{\Upsilon}{2}\right)^{-1},
\end{eqnarray}
where we have introduced $M_{-1} = M/0.1 M_{\odot}$.

The Schwarzschild criterion is given by \cite{schw,schw2}:
\begin{align*}
 \nabla_\text{rad}\leq&\nabla_\text{ad}\;\;\text{\small pure diffusive radiative or conductive transport}\\
  \nabla_\text{rad}>&\nabla_\text{ad}\;\;\text{\small adiabatic convection is present locally}.
\end{align*}
The adiabatic gradient $\nabla_\text{ad}$ depends on the properties of the gas. In the case of an ideal gas model, the adiabatic gradient is a constant, that is, $\nabla_\text{ad}=0.4$.

Considering a case when a fully convective star reaches the MS, that is, its interior conditions are sufficient to burn hydrogen in a stable way, one may find a maximal mass of a fully convective star on the MS. To do so, one needs to compare the luminosities of the hydrogen burning $L_H$ and the one responsible for the onset of the radiative core's development (we assume that opacity does not change); that is, when $\nabla_\text{rad}$ drops to $\nabla_\text{ad}$. The luminosity of hydrogen burning in Horndeski gravity was obtained in \cite{sak1,sak2}
\begin{equation} \label{hlum}
    \frac{L_H}{ 5.2\times 10^6 L_{\odot}} = \frac{\delta_{3/2}^{5.487} M_{-1}^{11.973} \eta^{10.15}}{\omega_{3/2} \gamma_{3/2}^{16.46}\left(1+ \frac{3\Upsilon}{2} \right)^{3/2}(\eta+ \alpha)^{16.46}}.
\end{equation} 
Writing $0.4=\nabla_{rad}$ and equaling it to \eqref{nabla} gives us the minimum luminosity for a star to develop a radiative core:
\begin{eqnarray}\label{minlum}
&& \frac{L_{min}}{5.2\times 10^6 L_{\odot}} = \\
& & 2.66 \times 10^{84} \frac{\mu^{6.8 }\delta_{3/2}^{1.86}G^6 (b c T_{\text{eff}})^{0.8} M_{-1}^{4.4}}{(k_B N_A)^{6.8}\xi_R^{8.66} (-\theta'(\xi_R))^{1.73} \kappa_0^{0.8}} \left(1+\frac{\Upsilon}{2} \right) , \nonumber
\end{eqnarray}
where we normalized it by $5.2\times 10^6 L_{\odot}$. Comparing the minimum \eqref{minlum} luminosity with the hydrogen burning luminosity \eqref{hlum} and solving for the mass, we obtain
{\small
\begin{eqnarray}\label{mmsm}
M_{-1}&=&1.41\times 10^{11}\dfrac{  \gamma_{3/2} ^{2.17} \mu ^{0.9}  \omega_{3/2} ^{0.13}( b c T_{\text{eff}})^{0.11}G^{0.79}(\alpha+\eta )^{2.17}}{(k_B  N_A)^{0.9} \delta ^{0.48} \xi ^{1.14} \left(-\theta' \right)^{0.23} \kappa _0^{0.11} \eta ^{1.34}} \nonumber \\
&\times&\left(1+\frac{\Upsilon}{2} \right)^{0.11} \left(1+\frac{3 \Upsilon}{2} \right)^{0.2}.
\end{eqnarray}
}

We can relate the modified mass $M^{\text{mod}}$ with the GR mass $M^{\text{GR}}$ as
{\small
\begin{eqnarray}
    M^{\text{mod}}&=& \left(\frac{\gamma_{3/2}}{\gamma_{3/2}^{\text{GR}}} \right)^{2.17} \left(\frac{\delta_{3/2}^{\text{GR}}}{\delta_{3/2}} \right)^{0.48}\left(\frac{\theta'^{\text{GR}}}{\theta'} \right)^{0.23}\left(\frac{\xi_R^{\text{GR}}}{\xi_R} \right)^{1.14} \nonumber \\ &\times& \left(\frac{\omega}{\omega^{\text{GR}}} \right)^{0.13}\left(1+\frac{\Upsilon}{2} \right)^{0.11} \left(1+\frac{3 \Upsilon}{2} \right)^{0.2} M^\text{GR}. 
\end{eqnarray}
}
For instance, the modified mass for $\Upsilon=-0.6$ and $\Upsilon=0.2$ are, respectively, $M^{\text{mod}}_{\Upsilon=-0.6}=0.7$ and $M^{\text{mod}}_{\Upsilon=0.6}=1.15$.

\section{Uncertainties analysis}
In what follows, we will investigate the uncertainties which our modelling can carry because of the crude assumptions. To do so, we will base this analysis on the one performed in \cite{saltas}.

To recall, we model the pre-Main Sequence star as a fully convective sphere of mass $M$ and radius $R$ with a thin layer of atmosphere at $r\approx R$. Therefore, the matter inside the star is well described by the polytropic EoS \eqref{pol}. The weakest part in our modelling is opacity whose relation is given by different forms of the Kramer law \eqref{abs}. Moreover, in our results we have used the constant values of the hydrogen and helium fractions, as well as metallicity.

Therefore, looking at our results, we clearly see that the main variables to be analysed are temperatures, opacities. and composition. As it will be clear from the following analysis, fractional changes of those variables are functions of the fractional change of density, mainly. Because of that fact, let us analyse the density solution for the modified density which was given in \cite{saltas} ($z:=r/R$): 
\begin{equation}
    \rho(z,\Upsilon) \simeq \rho_{(0)} + \epsilon\rho_{(\Upsilon)}(z),
\end{equation}
where $\epsilon\sim\Upsilon\ll1$ and $\rho_{(\Upsilon)}(r)$ is the contribution of scalar field to the density profile while $\rho_{(0)}(r)$ is the solution of the Newtonian hydrostatic equilibrium equation, given by, respectively, by the equations (14) and (10) in \cite{saltas}. Since the fractional change of density
\begin{equation}
    \frac{\delta\rho}{\rho} = \frac{\rho_{(0)} -\rho(z,\Upsilon)}{\rho_{(0)}}
\end{equation}
is singular for $z=1$, we may expand it around the surface to get 
\begin{equation}\label{frac}
    \frac{\delta \rho}{\rho} = -0.21 \Upsilon - 10.1\Upsilon\left(\frac{r-R_{cz}}{R_{\odot}}\right)+ ...,
\end{equation}
which is provided for the base of the convective zone being placed at $R_{cz}=0.99 R $. As demonstrated below, all crucial ingredients which can carry uncertainties are dependent on the above fractional change.

It is easy to see that the fractional change of temperature \eqref{temps} is
\begin{equation}
    \frac{\delta T}{T} =  \frac{\delta \mu}{\mu} +  \frac{1}{3}\frac{\delta \rho}{\rho},
\end{equation}
where $ \frac{\delta \mu}{\mu} = - \frac{\delta Z}{(5X-Z+3)}$  since we keep the hydrogen mass fraction $X = 0.7$ and helium $Y=1-Z-X$ constant, while the metallicity $Z$ varies from $0.001$ to $0.03$, and the mean molecular weight is
\begin{equation}
    \mu = \frac{4}{5X-Z+3}
\end{equation}
for a fully ionised gas. The fractional change of central temperature is given as
\begin{equation}
      \frac{\delta T_c}{T_c} =  \frac{\delta \mu}{\mu} +  \frac{2}{3}\frac{\delta \delta_n}{\delta_n},
\end{equation}\label{fractemp}
where $\delta_n$ is a numerical solution of the modified Lane-Emden equation, so it is also a function of \eqref{frac}.

Moreover, the fractional change of the $H^{-}$ opacity for the surface layer \eqref{hydro}
\begin{equation}
     \frac{\delta \kappa}{\kappa} \approx 9  \frac{\delta T}{T} +  \frac{1}{2}\frac{\delta \rho}{\rho} + \frac{\delta Z}{Z} \approx \frac{7}{2}\frac{\delta \rho}{\rho} + \frac{\delta Z}{Z} + 9 \frac{\delta \mu}{\mu}
\end{equation}
Therefore, the main uncertainty is related to the metallicity.

Let us now consider some reference values which we were using through the paper. For the solar metallicity $Z=0.02$, the fractional changes in $\mu$ are about $10^{-3}$ while in $Z$, when we vary it the solar value, they are about unity. On the other hand, the fractional change in density for $-0.6<\Upsilon < 0.391 $ \cite{rosyadi} are, respectively, $0.17$ and $0.12$ for the boundary limits. For  $-10^{-4}<\Upsilon < 5\times10^{-3} $ \cite{saltas} we have respectively $\sim-10^{-5}$ and $\sim-10^{-3}$. The fractional changes of $\delta_n$ for $\Upsilon = \pm 0.3 $ are of order $10^{-1}$ while for, for instance, for $\Upsilon=\pm10^{-4}$, it of order $10^{-4}$.

\section{Discussion and conclusions}

In this work we have studied the early evolution of low-mass stars within the framework of scalar-tensor extensions of gravity. As a working example, we have used the framework derived for Horndeski and beyond gravity, such that the modification to the hydrostatic equilibrium and other relevant equations are governed by the parameter $\Upsilon$. Consequently, several features of the early evolution of low-mass fully convective stars are modified, such as their temperatures at the photosphere, lithium abundances, and heat transfer processes. 

We start by investigating how the temperature--luminosity relation is altered in ST gravity. It was already demonstrated in various works \cite{aneta2,merce} that the Hayashi tracks are shifted in the case of modified gravity, therefore a similar result was also expected for non-zero values of $\Upsilon$. Our rough results - let us recall that the considered model lacks the appropriate atmosphere description, such that we do not get the realistic temperatures - are given in the figure \ref{fig}.
We see that the Hayashi track method can bound the positive values of the parameter $\Upsilon$, although the simplified model we have used do not allow to distinguish the curves which are given with the restrictive constraints obtained with the use of the helioseismic analysis \cite{saltas}. 

When a PMS star follows its Hayashi track, light elements, such as deuterium and lithium for instance, can be already burnt if the core conditions are sufficient to start the corresponding reactions. In this work we have focused mainly on the lithium one, since this result has serious consequences, as discussed in \cite{aneta3} and in the further part of the conclusions. Therefore, we proceed to calculate the lithium-to-hydrogen ratio as a function of time, given by the equation \eqref{ratio}, and this depletion rate could be also easily generalized to non-resonant reactions, which is given by \eqref{ratio2}. Those results can be subsequently fitted to the observational data,  {however, since most of the low-mass stars with masses $\approx 0.35 M_\odot> M>0.2M_\odot$ have consumed lithium just before reaching the Main Sequence (so we could deal with comparable destruction and contraction times)}, we could obtain the age, central temperature, radius, and luminosity of such a star, presented in the table \ref{tab_age}. It was done by equalling the lithium destruction time \eqref{destr} with the contracting time, that is, the time the stars stays on the Hayashi track \eqref{cont}, such that the resulting equation \eqref{central} could be solved together with \eqref{ctemp}, \eqref{Rpol}, and \eqref{stefan}. It is evident that those results are dependent on the theory of gravity, as observed already in \cite{aneta3}. We notice that although $T_c$, as well as radius and luminosity do not change too much with respect to GR values, one observes a significant change in the age. The positive values of $\Upsilon$ provide younger objects, however their biggest values can be probably discarded, since with increasing $\Upsilon$ we are approaching already the forbidden Hayashi zone. The negative values, on the other hand, provide that the star stays longer in the PMS phase such that it could happen than the whole evolution, from the proto-star to the white dwarf, would also be prolonged, providing that the age of very low-mass white dwarf stars would be greater than the age of the Universe \cite{masuda,lau}.

Finally, we have reexamined the Schwarzschild criterion which is also affected by the ST modifications, as evident from \eqref{schw}. This property, together with the previously derived luminosity of hydrogen burning in Horndeski gravity \cite{sak1,sak2}, allowed us to find the mass of a fully convective star on the MS \eqref{mmsm}. Similarly as for Hayashi track, we have used analytic opacity models, \eqref{bf} and \eqref{ff}, which does not provide realistic results. However, we can see from the mass ratios that the differences can be significant for higher values of the parameter $\Upsilon$.
Nevertheless, even a small difference will provide a slight change in the evolution of a particular star: in order to model its distinctive phases one uses the Schwarzschild criterion. When working in a theory of gravity which provides additional terms to the non-relativistic structure equations one should take those modifications into account in the modelling of the stellar evolution.

 When we examine the Hayashi expression \eqref{hay}, we can easily notice that the presence of the term $\left( 1+\frac{\Upsilon}{2}\right)^{\frac{4}{3}}$ will slightly increase or decrease $T_{ph}$ depending on whether $\Upsilon>0$ or  $\Upsilon<0$. However, this term is not the sole responsible for the shifting as $\theta'$ and $\xi_R$ also get modified as we change the parameter $\Upsilon$. The radius and luminosity depend on the age which, in turn, depends implicitly on $\Upsilon$ through $\xi_R$, $\theta'$ and $\delta$. The same applies to the central temperature, given by \eqref{central}. What we observe from table 1 is that, for negative values of $\Upsilon$, the central temperature decreases, while age, radius and luminosity increase. For positive values of $\Upsilon$, we have the opposite effect. Usually, negative values of $\Upsilon$ are related to an enhancement of gravity. Strengthening gravity means that other physical processes, as for example light elements' ignition in the stellar core can happen in lower temperatures than the ones obtained by assuming Newtonian model.

Let us now briefly discuss the consequences of our findings. 
As already discussed \cite{aneta3}, the light elements abundance at the photosphere, especially lithium, is an age-dependent quantity, allowing to determine young clusters’ age and individual stars \cite{basri,chab,bild,usho}. The lithium depletion method has been believed so far to be the most reliable technique for young globular clusters’ age determination and individual stars such as white dwarfs. Since it also depends on the theory of gravity in use, it can contribute to the explanation of “too old” white dwarfs by reducing the PMS phase by a few Myr \cite{aneta3}. Moreover, the different phases of the stellar evolution can be shorter or longer, therefore one deals with a different number of stars in the pre- and MS phases, giving diverse impact to the distant galaxies brightness in comparison to the GR prediction \cite{davis}. Subsequently, the lithium abundance may also be a tool to test theories of gravity: theories which prominently prolong the low-mass stars’ lifetimes in comparison to the current accepted models would rise doubts on those which introduce such an effect.

This finding has also something to say about the cosmological lithium problem, and the big bang nucleosynthesis (BBN) \cite{zyla}. Primordial nucleosynthesis is a probe of the very early universe as well as of the standard model physics and beyond, providing strong constraints on them. This is so because the light elements' abundances are sensitive to the few minutes old universe's conditions, thus different cosmological scenario, provided by modified gravity, influences their theoretical values \cite{bhat,ben,capN,nes}. These results are then compared to the observed abundances, that is, light element to hydrogen ratios which are estimated by observing halo stars of Milky Way. These stars are the oldest stars in our galaxy and they are believed to contain the primordial lithium in their atmospheres \cite{helmi}. It turns out that there is a significant discrepancy between the predicted and observed values in the lithium abundance, named as ``lithium problem" \cite{boe,fields}. However, as demonstrated in this paper and the previous one \cite{aneta3}, these ratios depend on gravitational theory which one uses to get the ratios. Therefore, in order to examine the lithium problem in cosmology, one should not only take into account the modified gravity effects affecting expansion rate used for obtaining the relic abundances of light elements, but one also should apply the considered modified gravity theory to lithium abundances in the halo stars. Currently, the common approach is to modify the BBN abundances only with respect to a particular theory of gravity, and then to compare them to observed ones, which were obtained by assuming Newtonian gravity. But, as already noticed, many theories of gravity modify the Newtonian physics, therefore this approach is not consistent. The works in these lines is in progress and will be presented somewhere else.

On the other hand, we have also discussed how modified gravity changes the scenario of the radiative core development. Therefore, if one studies an evolution of a given star in modified gravity, one should remember to take into account the effects on the convective imbalance. Moreover, if one is able to clearly distinguish low-mass stars on Henyey tracks \cite{hen1,hen2,hen3} from the ones following Hayashi ones, it could also be an opportunity to constraint a theory parameter: modified gravity also alters the maximum mass of a fully convective star on the MS, as discussed in the section \ref{radcore}.

Let us now discuss the uncertainties. We see that in the case of the temperature profile \eqref{fractemp} and its central values, the modified gravity effects are more important (they are two order of magnitude larger) that changes in the compositions for the less restrictive values given in \cite{rosyadi} while for the restrictive ones the composition is one order of magnitude larger than the contribution ruled by $\Upsilon$. On the other hand, in our modelling of the atmosphere, the main source of the uncertainties is the metallicity, dominating over composition and modified gravity. Definitively, without improving the atmosphere modelling to reduce this uncertainty, one cannot use the Hayashi tracks and fully convective stars on the Main Sequence to constrain this model of gravity. Apart from it, one should also improve the microphysics description such as an equation of state, electron degeneracy, phase transition points and ionization, to numerate just a few of them. Moreover, since it was demonstrated that the microscopic quantities are dependent on a given theory of gravity \cite{kim,del,hos1,hos2,awmicro,kalitaWD}, very likely the presence of the scalar field will also have a non-negligible effect to their forms and the processes happening in the stellar interiors. For further modelling, we should also consider rotating objects in modified gravity as it also influences the density profile and other physical properties \cite{chow2}. Moreover, as shown in \cite{auddy,kart}, the evolution of the electron degeneracy plays a crucial role in the contracting low-mass stars, therefore this fact should be also taken into account in the improved modelling. It is so because as demonstrated in this work as well as in \cite{maria,kart}, the modified gravity effects are more relevant for the older stellar and substellar objects, since those effects accumulates with time \cite{leane}.
 We will leave this analysis for the future work.

As a last comment, let us notice that in spite of the fact that modified gravity alters the stellar description on many different levels, there are not enough studies and methods developed for using stellar and substellar objects to test theories of gravity. As clearly demonstrated in \cite{baker}, there is a still not tested region corresponding to stars and galaxies (with the curvature-to-gravitational potential parametrization) between the problematic region - cosmology - and the regions in which GR provides a satisfactory description, that is, the Solar System and compact objects.

\section*{Acknowledgements}
   AW acknowledges financial support from MICINN (Spain) {\it Ayuda Juan de la Cierva - incorporac\'ion} 2020 No. IJC2020-044751-I.



\end{document}